\documentclass[aps,showpacs,twocolumn]{revtex4}
\usepackage{epsfig,amsmath,graphics}
\usepackage{graphicx}
\begin{document}
\title{Non-equilibrium Born-Oppenheimer potential energy surfaces for 
molecular wires}

\author{D.S. Kosov}      
\email{kosov@theochem.uni-frankfurt.de}

\affiliation{Institute of Physical and Theoretical Chemistry, 
J. W. Goethe University, Marie-Curie-Str. 11, D-60439 Frankfurt, Germany}
\begin{abstract}
We present a method for computing non-equilibrium, current-dependent
Born-Oppenheimer potential energy surfaces for molecular wires. 
Calculations are performed for polyacetylene wire described 
by tight-binding model with electron-phonon interactions.
We find that dimerization of the polyacetylene wire is 
amplified by electric current flow. We show that the
boundary between transparent and opaque black states
of the wire is blue-shifted by current.
\end{abstract}
\maketitle

\newpage
One of the major goals in modern nanotechnology is the construction
of an electronic circuit in which molecular systems act as conducting 
element~\cite{nitzan03}. Two critical issues should be thoroughly
investigated to proceed with the design of real molecular devices: heating 
in a molecular junction and current induced conformational changes of
a junction caused by electromigration. By the inelastic 
scattering of electrons the kinetic energy of current is transfered to
the nuclear vibrations and then it is released as heat.
The second process, electromigration, is the phenomenon of atom motion due 
to current induced forces and  it is related to the transfer of 
electronic momentum to nuclei due to
both elastic and inelastic scattering.
The electromigration can lead to  device breakdown at 
critical values of current and therefore current-induced forces 
may present a limitation for the development of molecular 
electronic devices. Very recently,
controlled electromigration has been used constructively 
to make a single-molecular junction \cite{liang02}. 
Both issues, heating and electromigration, are  ultimately related to 
understanding how 
Born-Oppenheimer (BO) potential energy surface of a molecular wire 
is responded to electronic current flow.
If  current-dependent BO potential energy 
surface of a molecular device is available, we can compute current-induced 
forces as the corresponding gradients as well as we 
can perform the normal modes analysis to obtain the current-dependent
vibrations.

Although the practical importance of a current-dependent 
BO surface can not be underestimated, some intricate theoretical 
questions  connected to 
the difficulty in defining a generic variational energy functional
for non-equilibrium steady state systems
\cite{todorov00,diventra00,emberly01}
can be also puzzled out. Theoretical work to date has failed 
to define or even to prove the  existence of the
BO potential energy surface for molecular wires with 
current flow.
The principle theoretical difficulty is
the treatment of the transport problem as an electron transmission 
(Landauer approach~\cite{landauer70}) which is non-variational from the outset.
A wire with current is by definition an open quantum system and 
as long as one works within a non-variational transport theory 
it is likely not possible 
to comprise a geometrical optimization of the nuclear 
degrees of freedom with electronic transport
calculations within a single theoretical approach.
There is the opinion  that 
current induced forces can not be conservative \cite{sorbello98}
although validity of this claim has been recently
questioned \cite{diventra03}.  
If a current 
induced force is not conservative then  there does not exist a 
BO potential energy surface of which this force might be 
derivative.
One of the goals of this Letter is to show  that this is not true and that 
the Born-Oppenheimer 
potential energy surface can be properly defined and efficiently computed 
in the presence of  current flow.

The Lagrange multiplier based transport theory~\cite{kosov02,kosov03-jcp}
takes its origin in the  modern development of steady state non-equilibrium
statistical mechanics~\cite{antal97,antal99,eisler03}. This theory 
is equivalent to a variational formulation where the constrained 
minimization of the expectation value of the Hamiltonian on the manifold of 
the desired current  $J$
yields the transport properties~\cite{kosov02}. 
The ``algorithm'' of the Lagrange multiplier based transport theory 
to compute a current-dependent BO surface  
can be summarized as follows:
(1) define the operator of the current  $J$ via the continuity equation;
(2) extend the Hamiltonian $H$ by adding the term 
($ -\lambda J$) where $\lambda$
is a Lagrange multiplier; 
(3) compute the BO surface as the expectation value  
$\langle 0(\lambda)|H| 0(\lambda) \rangle $ where 
$ |0(\lambda) \rangle$ is the ``ground state'' of the extended Hamiltonian
$H -\lambda J$. 

In this Letter, using the Lagrange multiplier based transport theory
we compute the current-dependent BO potential energy surfaces 
of  conjugated polymer, polyacetylene, wire.
Polyacetylene is a linear polymer, it consists of coupled chain
of $CH$ units forming a quasi-one-dimensional lattice. Each $C$ has 
four valence electrons: 
two of them contribute to the $\sigma$-type bonds 
connecting neighboring carbons along
the one-dimensional backbone, while the third forms a bond with the 
the hydrogen. The remaining electron has the symmetry of a $2p_z$
orbital and contributes to $\pi$-bond. In terms of the energy 
band description, the $\sigma$-bonds form low-lying completely
filled band, while $\pi$-bond leads to half-filled energy band 
responsible for electron transport properties. The  ${\pi}$-electrons can to 
a fair approximation be treated as 
a quasi-one dimensional electron gas within the tight-binding 
approximation.  Accordingly, we start our consideration from the 
following model electronic Hamiltonian for the polymer chain
\begin{eqnarray}
H&=& -\sum_{n\sigma}({t_o}+(-1)^n \alpha x)
(a_{n+1\sigma}^{\dagger} a_{n\sigma}
+a_{n\sigma}^{\dagger} a_{n+1\sigma})
\nonumber
\\
&+& 2NKx^2 \;. 
\label{h0}
\end{eqnarray}
Here $a_{n \sigma}$ creates an electron of spin $\sigma$ on site $n$.
The Hamiltonian $H$ is based upon the so called
Su-Schrieffer-Heeger (SSH) \cite{heeger88} model for the band-structure of
{\em trans}-(CH)$_n$ lattices.  The first term in the $H$ gives the
energy for an electron with spin $\sigma$ to hop between neighboring
$p_z$-orbitals. The strength of this term is modulated by linear
coupling to distortions $x$ in the polymer lattice away from evenly spaced
lattice positions. Finally, the last terms in $H$ gives the harmonic
interactions between lattice-sites arising from the $\sigma$-bonds
between neighboring lattice atoms. 

We begin the derivation by defining the operator of current via 
continuity equation. 
The number of electrons on the site $n$ is given by the
expectation value of the
operator 
\begin{equation}
N_n = \sum_{\sigma} a^{\dagger}_{n \sigma}  a_{n \sigma}\;.
\end{equation}
By making use of the Heisenberg representation
 the continuity equation can be written as 
the Heisenberg equation-of-motion for the operator $N_n$:
\begin{equation}
\dot{N_n} = i \left[ H_o, N_n \right] \; .
\label{cont-eq-1}
\end{equation}
Given standard anti-commutation relations between the electron 
creation and annihilation operators, the  r.h.s. commutator 
(\ref{cont-eq-1}) can be readily computed.
Comparing  eq.(\ref{cont-eq-1}) with the finite difference 
expression for continuity equation
$\dot{N_n} = -(j_n -j_{n-1})$  
we obtain the definition of the operator of  current through  site $n$.
By making the sum of  on-site currents $j_n$ along the wire we define 
the net, macroscopic current through the wire
\begin{equation}
J = i \sum_{n \sigma} ({t_o}+(-1)^n \alpha x) 
(a^{\dagger}_{n+1 \sigma} a_{n \sigma} - 
a^{\dagger}_{n \sigma} a_{n+1 \sigma})\;.
\label{current}
\end{equation}
There are two distinct contributions to the net current. The term 
${t_o}(a^{\dagger}_{n+1 \sigma} a_{n \sigma}-
a^{\dagger}_{n \sigma} a_{n+1 \sigma})$ is the standard for 
molecular wires without electron-phonon coupling whereas 
the term  $(-1)^n \alpha x (a^{\dagger}_{n+1 \sigma} a_{n \sigma} - 
a^{\dagger}_{n \sigma} a_{n+1 \sigma})$ is the phonon-assisted current.

We assume now that there is a 
time-independent, constant current through the wire.
A homogeneous  current-carrying state is the 
same whether it is introduced by  reservoirs
or by a bulk driving driving field, the current enters the
problem via a
Lagrange multiplier $\lambda$.  To this end the Hamiltonian $H$  
is modified by adding the term which constraints the macroscopic 
current $J$:
\begin{equation}
H_J=H - \lambda J\;.
\label{h0-lambdaj} 
\end{equation}
The Hamiltonian $H_J$ is  a quadratic form in 
fermion creation/annihilation operators. This quadratic form
can be exactly diagonalized by two unitary transformations.
The  Fourier transformation
\begin{equation}
a_{n \sigma} = \frac{1}{\sqrt{N}} \sum_{k} \exp(ikn) c_{k \sigma}, 
\end{equation}
brings the Hamiltonian (\ref{h0-lambdaj}) to the following form
\begin{eqnarray} 
H_J=\sum_{0\le k<2\pi} \sum_{\sigma} 
\left( c^{\dagger}_{k+\pi \sigma} c^{\dagger}_{k \sigma} \right)
\left(
\begin{array}{cc}
\epsilon_k & \Delta_k \\
\Delta_k & \epsilon_k
\end{array}
\right)
\left(
\begin{array}{c}
c_{k+\pi \sigma}\\ 
c_{k \sigma}
\end{array}
\right).
\label{h1}
\end{eqnarray}
We have introduced here the phonon-unperturbed
band energy
\begin{equation}
\epsilon_k(\lambda) = 2t_0 (\cos(k) + \lambda \sin(k)), 
\end{equation}
and the phonon-induced gap
\begin{equation}
\Delta_k (\lambda,x) = -4 \alpha x (\sin(k) - \lambda \cos(k)). 
\end{equation}
Both quantities, $\epsilon_k(\lambda)$ and  $\Delta_k (\lambda,x)$
depend upon current via the Lagrange multiplier $\lambda$
and if we let $\lambda$ tend to zero we recover the usual zero-current 
results~\cite{heeger88}.
Mixing $c_{k \sigma} $ and $c_{k+\pi \sigma}$ operators
by the canonical  Bogoliubov transformation 
\begin{eqnarray}
\left(
\begin{array}{c}
\alpha_{k+\pi \sigma}
\\
\alpha_{k \sigma}
\end{array}
\right)
=
\left(
\begin{array}{cc}
u_k
&
v_k
\\
v_k
&
-u_k
\end{array}
\right)
\left(
\begin{array}{c}
c_{k+\pi \sigma}
\\
c_{k \sigma}
\end{array}
\right),
\end{eqnarray}
we obtain new
quasipaticles $ \alpha_{k \sigma} $ and $\alpha_{k+\pi \sigma}$.
The coefficients of the Bogoliubov transformation is
chosen in such a way that the extended Hamiltonian (\ref{h1}) should be
diagonal. It leads to the following expressions for the coefficients 
\begin{equation}
u_k = \sqrt{\theta(\epsilon_k)}\cos(\phi_k) 
-\sqrt{1-\theta(\epsilon_k)}\sin(\phi_k)
\;,
\end{equation}
\begin{equation}
v_k=\sqrt{\theta(\epsilon_k)}\sin(\phi_k) +
\sqrt{1-\theta(\epsilon_k)}\cos(\phi_k)\;,
\end{equation}
where $\theta(\epsilon_k)$ is the Heaveside step function and the 
mixing angle $\phi_k$ is given by the
formula $\phi_k=1/2 \arctan(\epsilon_k/\Delta_k)$.
This canonical Boguluibov transformation produces the non-interacting 
Hamiltonian for the quasiparticles 
$ \alpha_{k \sigma} $ and $\alpha_{k+\pi \sigma}$:
\begin{eqnarray}
H_J&=&\sum_{0 \le k<\pi} \sum_{\sigma} E_k(
\alpha^{\dagger}_{k+\pi \sigma} \alpha_{k+\pi\sigma} 
-\alpha^{\dagger}_{k\sigma} \alpha_{k\sigma} )
\nonumber
\\
&+& 2NKx^2 \;,
\label{hdiag}
\end{eqnarray}
 with 
the wave-vector $k$ in the reduced Brillouin zone of the system.
The band energy, $E_k=\sqrt{\epsilon_{k}^{2}+\Delta_{k}^{2}}$, 
and the mixing angle $\phi_k$ in the canonical Bogoliubov transformation
are functions of 
the phonon-unperturbed band energy, $\epsilon_k(\lambda)$, and the 
phonon-induced energy gap, $\Delta_k(\lambda,x)$. 

It can be straitforwardly shown that this diagonalization 
of the Hamiltonian is equivalent to the variational problem 
$ \delta \langle H -\lambda J \rangle =0$ with the coefficients of
the Bogoliubov transformation $u_k$ and $v_k$ as variational parameters.
Current  carrying steady states are by definition the ground states of
extended, current dependent Hamiltonian $H_J$ \cite{kosov02} .
The ground state of the Hamiltonian (\ref{hdiag}) can be defined as the 
vacuum state for quasiparticle $\alpha_{k+\pi \sigma}$
and $\alpha^{\dagger}_{k\sigma}$.
All physical properties of the system depend upon the 
current and can be obtained as the matrix element of the corresponding 
operators over the current carrying ``ground state'' $|0(\lambda)\rangle $. 
Now the non-equilibrium Born-Oppenheimer potential energy 
surface is defined as 
the expectation value of the original Hamiltonian $H$ (\ref{h0}) over the 
current carrying wave function:
\begin{eqnarray}
&&E_{BO}(\lambda, x) =\langle 0(\lambda)|H|0(\lambda)\rangle 
\nonumber
\\
&&= 
-2\sum_{0 \le k<\pi} E_k(\lambda,x) + 2NKx^2
\end{eqnarray}
\begin{figure}
  \centerline{
\epsfig{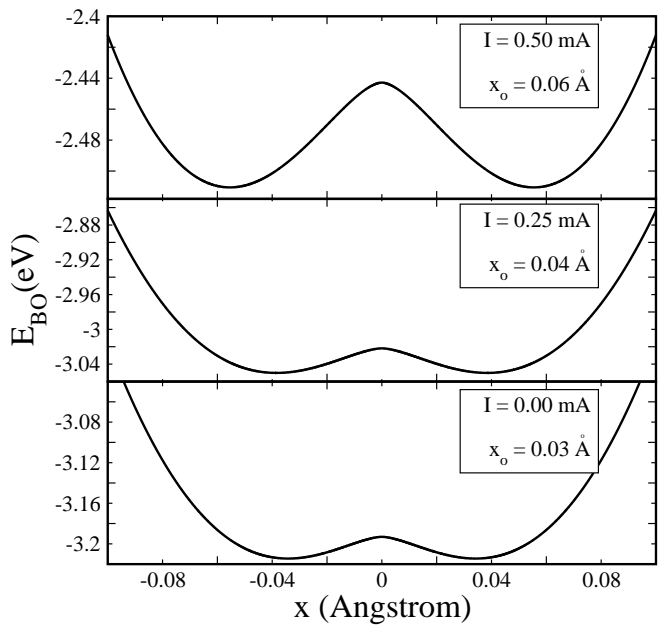}}
\caption{Born-Oppenheimer potential energy surfaces for different values
of the current: I=0.00 mA (lower panel), I=0.25 mA (middle panel)
and I=0.50 mA (upper panel). The dimerization coordinate, $x_o$,
corresponds to the minimum on the BO surface and it is shown 
for each values of the current.  
  }
\label{fig:bo}
\end{figure}
The  band energy $E_k$, the mixing angles $\phi_k$, and  the 
current carrying wave function $|0(\lambda)\rangle$ are not yet in a form 
to be computed as the Lagrange multiplier  
$\lambda$ is not known yet.
The additional equation for the Lagrangian multiplier
$\lambda$ is obtained if  
the density of an
expectation value of the net current operator (\ref{current}) 
over the current carrying state  of the
Hamiltonian (\ref{hdiag}) is required to yield a desired current density $I$.
This sets up the continuity equation:
\begin{widetext}
\begin{equation}
\frac{1}{N}\langle 0(\lambda)|J|0(\lambda)\rangle =
 \frac{2}{N}\sum_{0 \le k<\pi} 
(2 \theta(\varepsilon_k) -1 ) \left( 2 t_0 \sin(k) \cos(2\phi_k) +
4 \alpha x   \cos(k) \sin(2\phi_k) \right)=I
\label{I}
\end{equation}
\end{widetext}
For any given lattice distortion  $x$ the nonlinear equation (\ref{I})
should be resolved for $\lambda$. Then the band energy and the mixing angle
$\phi_k$ of the canonical Bogoliubov transformations are computed as 
functions of current.

We now discuss the main results obtained using the above model.
The numerical calculations are performed for 
the half-occupied conductance band with the box 
length $N=1000$. In all our calculations we use the
following model parameters of the SSH model ~\cite{ness01}:
$t_o=2.5$ eV, $\alpha=4.1$ eV, and $K=21$ $eV/\overcirc{A}^2$.
This parameterization of the ASH Hamiltonian is typically used 
to describe coherent electron transport in polyacetylene wire~\cite{ness01}.

\begin{figure}
  \centerline{
\epsfig{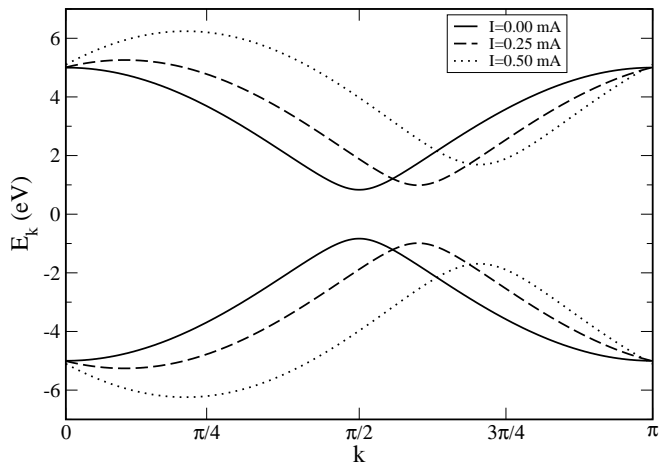}}
\caption{The current-dependent valence and conductance bands 
of polyacetylene wire,
$E_k= \pm \sqrt{\epsilon_{k}^{2}+\Delta_{k}^{2}}$.}
\label{fig:band}
\end{figure}

In the first calculations, shown in Fig.1, we studied the changes of the
BO surface as current flows through the wire.
 It is immediately evident from Fig.1 that the 
current substantially alters the Born-Oppenheimer potential 
energy surface of the polyacetylene wire. 
The minima on the BO surface correspond to  the optimal non-equilibrium 
steady state geometries of the wire. The dimerization coordinate
increases as current flows, i.e. the $C=C$ double bonds 
are shortened 
and the $C-C$ single bonds are elongated by the current induced forces. 
By increasing the current we also increase the barrier between the 
two-fold degenerate ground states,
therefore the energy difference between evenly spaced polyacetylene and
dimerized  polyacetylene is increased as we increase the current.
Since SSH Hamiltonian is very
general and applicable in its present formulation to any long semiconductor
wires we do expect that effect of the current-amplified 
dimerization occurs in other quasi-one-dimensional wires as well.
The results of our calculations are qualitatively consistent 
with the finding of Todorov, Hoekstra and Sutton \cite{todorov01} 
that current-induced forces generally form an alternating patterns in
atomic wires.
The critical value of the current at which the current induced
forces break the wire into pieces can be also determined.
If we assume that the
current-induced bond distortion larger than 0.2$\overcirc{A}$
will be sufficient to rupture the wire, 
the critical value of the current which the polyacetylene wire can sustain
is 0.8 mA.

Finally, we turn our attention towards 
what physical quantities which are accessible experimentally
can be used to monitor  the current-amplified 
polyacetylene dimerization. To this end we compute the response of the
band structure of polyacetylene to current flow. 
The current-dependent band structure of the first Brillouin zone
is plotted in Fig.2. Polyacetylene's band is half occupied 
therefore the states with negative $E_k$ are completely 
occupied and the states with positive $E_k$
are empty.  The current flow through the wire does not 
only produce the overall shift of the band energy within the 
reduced Brillouin zone but also changes the the total bandwidth 
from 10 eV at zero current to 12.5 eV at $I=0.5$ mA. The band gap 
between the valence and conductance bands is also modulated by current. 

\begin{figure}
 \centerline{
\epsfig{figure=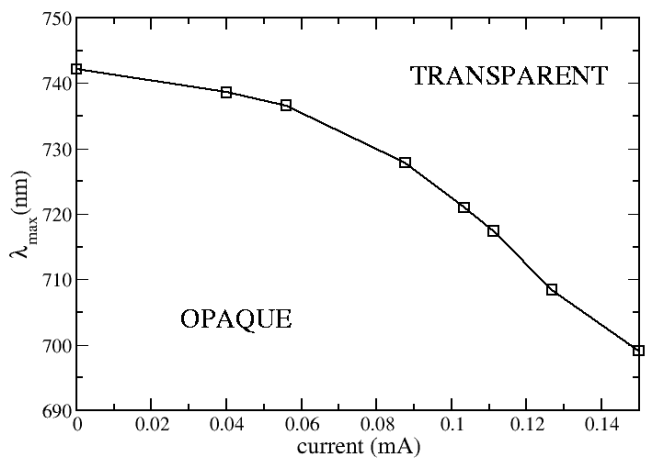,
width=1\columnwidth}}
\caption{The transparent-opaque boundary for polyacetylene wire as
a function of current}
\label{fig:i-wavelength}
\end{figure}
 
From the band structure calculations  
the band gap can be extracted as a function of the current.
Then the band gap can be turned into the  longest possible  wavelength of 
absorbed radiation $\lambda_{max}$ and
this is plotted in Fig.3 for different 
values of the net current density. 
Being a semiconductor wire, polyacetylene 
is transparent to the photons with wave 
length longer than $\lambda_{max}$ (i.e. with the energies within
the band gap)
but it is opaque to photons  with wavelength shorter than $\lambda_{max}$
(within the bandwidth).
It is clear from that the boundary between transparent and opaque
black parts of the spectrum
is blue-shifted by current flow in the polyacetylene wire. In other words, the 
color of the polyacetylene wire can be modified by electric current.   

In this Letter, we have laid a theoretical ground-work for computing a 
Born-Oppenheimer potential energy surface for molecular wires with 
direct current. Within the Lagrange multiplier based transport theory
we presented a computational tractable theoretical scheme to compute BO
of molecular wire junctions. Based upon applications of the techniques to 
tight-binding model with electron-phonon interactions we predict
that 
(1)the dimerization of the polyacetylene wire is amplified by current;
(2)the transparent-opaque boundary is blue-shifted by current flow;
(3)the generic mechanism of the semiconductor wire breakage is
 current increased dimerization.

The author wishes to thank A.Nitzan for valuable discussion.



\begin{thebibliography}{50}

\bibitem{nitzan03} A. Nitzan and M.A. Ratner,
Science {\bf 300}, 1384 (2003).


\bibitem{liang02} W.J. Liang, M.P. Shores, M.Bockrath, J.R.Long, and 
H.Park, Nature {\bf 417}, 725 (2002).

\bibitem{todorov00} T.N. Todorov, J.Hoekstra and A.P. Sutton,
Phylos. Mag. B {\bf 80}, 421 (2000)

\bibitem{diventra00} M. Di Ventra and S.T. Pantelides, Phys.Rev.B
{\bf 61}, 16207 (2000).

\bibitem{emberly01} E.G.~Emberly and G.~Kirzenow, Phys.Rev.B {\bf 64},
125318 (2001)

\bibitem{landauer70} R.Landauer, Philos. Mag. {\bf 21}, 863 (1970).

\bibitem{sorbello98} R.S. Sorbello, Solid State Physics 
(Academic Press,New York, 1998), Vol.51.

\bibitem{diventra03} M. Di Ventra, Yu-Chang Chen, T.N. Todorov,
cond-mat/0307739

\bibitem{kosov02} D.S.~Kosov, J.Chem.Phys. {\bf 116}, 6368 (2002).

\bibitem{kosov03-jcp} D.S.~Kosov, J.Chem.Phys., submitted (cond-mat/0309583)

\bibitem{antal97} T.Antal, Z.R\'acz, and L.Sasv\'ari,
Phys.Rev.Lett. {\bf 78}, 167 (1997).

\bibitem{antal99} T.Antal, Z.R\'acz, A.R\'akos, and G.M.Sch\"utz
Phys.Rev. E, {\bf 59}, 4912 (1999).

\bibitem{eisler03} V.Eisler, Z.R\'acz, and F. van Wijland, 
Phys.Rev. E, {\bf 67}, 056129 (2003).

\bibitem{heeger88}A. J. Heeger, S. Kivelson, J. R. Schrieffer, and W.-P. Su, 
Rev.Mod.Phys. {\bf 60},  781 (1988). 

\bibitem{ness01} H. Ness, S. A. Shevlin, and A. J. Fisher,
Phys. Rev. B 63, 125422 (2001). 

\bibitem{todorov01} T.N.~Todorov, J.~Hoekstra, and A.P.~Sutton,
Phys.Rev.Lett. {\bf 86}, 3606 (2001).


\end{thebibliography}
\end{document}